\documentclass[
12pt,
preprint,preprintnumbers,nofootinbib,
groupedaddress,superscriptaddress,amsmath,amssymb]{revtex4}
\usepackage{graphicx}
\usepackage{dcolumn}
\usepackage{bm}
\usepackage{amssymb}
\usepackage{amsmath}
\usepackage{epsfig}    
\usepackage{color}
\usepackage{hhline}

\def\be{\begin{equation}}
\def\ee{\end{equation}}
\newcommand{\bea}{\begin{eqnarray}}
\newcommand{\eea}{\end{eqnarray}}
\newcommand{\nn}{\nonumber}

\numberwithin{equation}{section}

\begin{document}
{\begin{flushright}{KIAS-P19023,  APCTP Pre2019 - 007}\end{flushright}}

\title{A modular $A_4$ symmetric model of dark matter and neutrino}

\author{Takaaki Nomura}
\email{nomura@kias.re.kr}
\affiliation{School of Physics, KIAS, Seoul 02455, Korea}

\author{Hiroshi Okada}
\email{hiroshi.okada@apctp.org}
\affiliation{Asia Pacific Center for Theoretical Physics (APCTP) - Headquarters San 31, Hyoja-dong,
Nam-gu, Pohang 790-784, Korea}

\date{\today}

\begin{abstract}
We propose a model based on modular $A_4$ symmetry containing a dark matter candidate, realizing radiatively induced neutrino mass at one-loop level. One finds that stability of dark matter candidate can be assured by nonzero value of modular weight and heavy neutral fermion mass hierarchies, which include dark matter under the $A_4$ triplet, are uniquely determined; $M_X\ll M_2<M_3$. Therefore we clearly identify single dark matter field which can discriminate from the other models with $A_4$ modular symmetry.  Then we discuss several phenomenological aspects and show predictions on the lepton sector.Especially, we find $0.56\lesssim\sin^2\theta_{23}\lesssim0.624$, which could have an advantage of this narrow region and be also discriminated from the other modular $A_4$ models.
\end{abstract}
\maketitle
\newpage

\section{Introduction}

The standard model (SM) of particle physics has been successfully confirmed by the experimental data including the discovery of the Higgs boson at the Large Hadron Collider (LHC). 
However physics beyond the SM is also indicated by some issues such as existence of dark matter (DM), non-zero tiny neutrino masses and origin of flavor structure.
In describing physics beyond the SM, symmetry would be a key aspect as the SM is based on the gauge symmetry.
In fact an additional symmetry such as discrete $Z_2$ can guarantee stability of DM and forbid neutrino mass generation at tree level, 
and it is often used in construction of radiative seesaw model~\cite{Ma:2006km}.
Furthermore non-abelian discrete symmetries have been applied to explain flavor structure in the SM~\cite{Altarelli:2010gt, Ishimori:2010au, Ishimori:2012zz, Hernandez:2012ra, King:2013eh, King:2014nza, King:2017guk, Petcov:2017ggy}.

Recently an interesting framework of symmetry has been considered in which modular group is applied and non-abelian discrete symmetries 
are obtained as their subgroups~\cite{deAdelhartToorop:2011re, Feruglio:2017spp, Baur:2019kwi}.
Then some works have been done applying the framework to flavor structure of leptons and/or quarks in $S_3$~\cite{Kobayashi:2018vbk, Kobayashi:2018bff}, $S_4$~\cite{Novichkov:2018ovf, Penedo:2018nmg, Kobayashi:2018bff}, $A_4$~\cite{Kobayashi:2018vbk, Kobayashi:2018scp, Criado:2018thu, deAnda:2018ecu, Okada:2018yrn, Novichkov:2018yse}, and $A_5$~\cite{Novichkov:2018nkm, Ding:2019xna}. 
One interesting feature of the framework is that couplings can be written as modular forms which are functions of modulus $\tau$ and transform non-trivially under the modular group.
Then we can use such a non-trivial structure of couplings to restrict interactions in a model.
Furthermore we do not need many scalar fields to break non-abelian discrete symmetries called flavon, breaking the symmetry by vacuum expectation value (VEV) of modulus $\tau$.

In this letter, we apply the framework of modular $A_4$ symmetry to construct a radiative seesaw model including a DM candidate.
The right-handed neutrinos are introduced as an $A_4$ triplet with modular weight $-1$.
We also introduce some scalar fields such as an isospin doublet field and the SM singlet fields with non-zero modular weight to realize neutrino mass generation at one loop level.
Interestingly the assignment of modular weight plays a role of a $Z_2$ symmetry in realizing radiative seesaw mechanism and guaranteeing stability of DM.
Moreover we obtain characteristic mass relation among three right-handed neutrinos since it is assigned as triplet under modular $A_4$.
We discuss phenomenologies of our model such as mass spectrum of right-handed neutrinos, relic density of DM and predictions on the lepton sector.

This letter is organized as follows.
In Sec.~\ref{sec:realization},   we explain our model setup under the modular $A_4$ symmetry.
Then, we discuss the right-handed neutrino mass spectrum, lepton flavor violation (LFV), relic density of DM and generation of the active neutrino mass at one loop level.
Finally we conclude and discuss in Sec.~\ref{sec:conclusion}.

\begin{center} 
\begin{table}[tb]
\begin{tabular}{|c||c|c|c|c|c|c|c||c|c|c|c|c||}\hline\hline  
&\multicolumn{7}{c||}{ Fermions} & \multicolumn{4}{c||}{Bosons} \\\hline
  & ~$L_{L_e}$~& ~$L_{L_\mu}$~ & ~$L_{L_\tau}$~& ~$e_{R}$~& ~$\mu_{R}$~ & ~$\tau_{R}$~ & ~$N_{R_a}$~  & ~$H$~& ~$\eta$~ &~$S$~ & ~$\varphi$~
  \\\hline 
 $SU(2)_L$ & $\bm{2}$  & $\bm{2}$  & $\bm{2}$ & $\bm{1}$   & $\bm{1}$  & $\bm{1}$ & $\bm{1}$ & $\bm{2}$   & $\bm{2}$ & $\bm{1}$  & $\bm{1}$   \\\hline 
$U(1)_Y$ & $-\frac12$ & $-\frac12$ & $-\frac12$  & $-1$& $-1$ & $-1$  & $0$  & $\frac12$   & $\frac12$ & $0$ & $0$   \\\hline
 $A_4$ & $1$ & $1'$ & $1''$ & $1$ & $1'$ & $1''$ & $3$ & $1$ & $1$ & $1$   & $1$    \\\hline
 $k$ & $0$ & $0$ & $0$ & $0$ & $0$ & $0$ & $-1$ & $0$ & $-1$ & $-3$  & $-2$    \\\hline
\end{tabular}
\caption{Field contents of fermions and bosons
and their charge assignments under $SU(2)_L\times U(1)_Y\times A_{4}$ in the lepton and boson sector, where $k$ is the number of modular weight, $a=1,2,3$, and the quark sector is the same as the SM.}
\label{tab:fields}
\end{table}
\end{center}

\begin{center} 
\begin{table}[tb]
\begin{tabular}{|c||c|c||c|c|c|c|c|c|c|c||}\hline\hline  
 &\multicolumn{2}{c||}{ Lepton couplings} & \multicolumn{7}{c||}{Higgs terms} \\\hline
  & ~$y_{\eta_i}(y'_{\eta_{i}})$~& ~$M_i$~ & ~$\mu_{S}$~& ~$\mu_{\varphi}$~& ~$\mu_{\varphi\eta}$~ & ~$\mu_{H\eta S}$~  & ~$\lambda_{\eta}^{(')}$~  & ~$\lambda_{S}$~  & ~$\lambda_{\varphi}$~  \\\hline 
 $A_4$ & $3(1,1',1'')$ & $3$ & $1$& $1$& $1$& $1$ & $1$& $1$& $1$     \\\hline
 $k$ & $2(4)$ & $2$ & $6$ & $4$ & $4$ & $4$&  $4$ & $12$  & $8$  \\\hline
\end{tabular}
\caption{Modular weight assignments for Yukawa and Higgs couplings, the other couplings are all neutral under the modular weight,
where $i=1,3$ denotes the component of $A_4$ triplet.
Notice here that the number of modular  weight for Higgs terms has to start at 4 because they are singlets under $A_4$ group.}
\label{tab:couplings}
\end{table}
\end{center}

\section{ Model} 
\label{sec:realization}
Here we explain our model with modular $A_4$ symmetry where some fields have non-zero modular weight and couplings with non-zero modular weight are modular forms.
First of all, we introduce three right-handed neutrinos as a triplet under $A_4$ and with modular weight $-1$,
where all the SM leptons have zero modular weight and assigned three kinds of singlet ${1,1',1''}$ for each flavor under the $A_4$ group. 
In the scalar sector, we introduce an isospin doublet field and two singlet fields $(\eta, \varphi, S)$ having non-zero modular weight $(-1,-2,-3)$,
where all the scalar fields are $A_4$ true singlet, $1$, and we assume $S$ to be real for simplicity. 
We expect that $S$ and $\eta$ are inert scalars in order to forbid tree level light active neutrino masses.
Each VEV of $H$ and $\varphi$ is denoted by $v_H/\sqrt2$ and $v_\varphi/\sqrt2$, where $H$ is identified as SM-like Higgs.
$\varphi$ plays a role in inducing the appropriate mass of $\eta$, as we will discuss later.
We summarize fields assignments in table \ref{tab:fields} and couplings in table \ref{tab:couplings}.
{\it In general remarks of couplings, we remind that any couplings with $A_4$ singlets $1,1',1''$ have to start from the number of modular wight $k=4,6,\cdots$, while any couplings with $A_4$ triplet start from $k=2,4,6,\cdots$. }
{This fact is understood as follows; modular form with $k=2$ is possible only when the associated coupling is triplet under $A_4$ and a coupling with $1, 1', 1''$ can be constructed by a tensor products of $k=2$  and $A_4$ triplet ones (pure singlet $1$ can be also considered as usual coupling constant which is not modular form without non-zero modular weight). 
In the framework, an interaction is invariant under modular symmetry when sum of modular weights for each associated fields and coupling is zero and it is invariant under $A_4$ symmetry. 
Then, interestingly, modular weight can play a role of $Z_2$ symmetry restricting interactions and stabilizing DM since any coupling should have even weight and odd number of odd weight fields in an interaction is forbidden. 
Thus even/odd modular weight correspond to $Z_2$ even/odd for the fields in the model.
}

Under these symmetries, one writes renormalizable Lagrangian as follows:
\begin{align}
-{\cal L}_{Lepton} &=
\sum_{\ell=e,\mu,\tau} y_{\ell}\bar L_{L_\ell} \ell_R H\nn\\
&+\lambda_{1}\bar L_{L_e} (y_{\eta_1} N_{R_1} + y_{\eta_2} N_{R_2}+ y_{\eta_3} N_{R_3} )\tilde\eta
\nn\\
&+\lambda_{2}\bar L_{L_\mu} (y_{\eta_1} N_{R_1} +\omega y_{\eta_2} N_{R_2}+\omega^2 y_{\eta_3} N_{R_3} )\tilde\eta
\nn\\
&+\lambda_{3}\bar L_{L_\tau} (y_{\eta_1} N_{R_1} +\omega^2 y_{\eta_2} N_{R_2}+\omega y_{\eta_3} N_{R_3} )\tilde\eta
\nn\\
&+ M_{N_3} (\bar N^C_{R_1} N_{R_2} + \bar N^C_{R_2} N_{R_1})
+ M_{N_2} (\bar N^C_{R_3} N_{R_1} + \bar N^C_{R_1} N_{R_3}) \nonumber \\
& + M_{N_1} (\bar N^C_{R_2} N_{R_3} + \bar N^C_{R_3} N_{R_2})
+ {\rm h.c.}, \label{eq:lag-lep}
\end{align}
where $\omega=e^{i\frac{2}{3}\pi}$ and the charged-lepton mass eigenstate is directly given by the first term above.
Thus, the observed mixing matrix for lepton sector is found in the neutrino sector only. 

In order to reproduce the neutrino oscillation data, we have to extend the Majorana mass sector so that
 nonzero diagonal elements are switched on. The way of this modification is carried out by introducing Yukawa couplings with
$(1,1',1'')$ and 4 under $A_4$ and modular weight assignments. Then the following Yukawa terms are added into the above Lagrangian:
\begin{align}
{\mathcal L}_{N N \varphi} = &\zeta_1 y'_{\eta_1}\varphi (\bar N^C_{R_1} N_{R_1}+\bar N^C_{R_2} N_{R_2}+\bar N^C_{R_3} N_{R_3})\nn\\
&\zeta_2 y'_{\eta_{2}}\varphi (\bar N^C_{R_1} N_{R_1}+\omega^2\bar N^C_{R_2} N_{R_2}+\omega\bar N^C_{R_3} N_{R_3})\nn\\
&\zeta_3 y'_{\eta_{3}}\varphi (\bar N^C_{R_1} N_{R_1}+\omega\bar N^C_{R_2} N_{R_2}+\omega^2\bar N^C_{R_3} N_{R_3})+{\rm h.c.},
\end{align}
where $y'_{1,2,3}$ is determined by modulus $\tau$ only as can be seen below, while $\zeta_{1,2,3}$ are free parameters.
After $\varphi$ developing VEV, we will find the diagonal elements of $M_N$ and reproduce the neutrino oscillation
data. 

The  modular forms of weight 2 $(y_{\eta_1},y_{\eta_2},y_{\eta_3})$ transforming
as a triplet of $A_4$ is written in terms of Dedekind eta-function  $\eta(\tau)$ and its derivative \cite{Feruglio:2017spp}:
\begin{eqnarray} 
\label{eq:Y-A4}
y_{\eta_1}(\tau) &=& \frac{i}{2\pi}\left( \frac{\eta'(\tau/3)}{\eta(\tau/3)}  +\frac{\eta'((\tau +1)/3)}{\eta((\tau+1)/3)}  
+\frac{\eta'((\tau +2)/3)}{\eta((\tau+2)/3)} - \frac{27\eta'(3\tau)}{\eta(3\tau)}  \right), \nonumber \\
y_{\eta_2}(\tau) &=& \frac{-i}{\pi}\left( \frac{\eta'(\tau/3)}{\eta(\tau/3)}  +\omega^2\frac{\eta'((\tau +1)/3)}{\eta((\tau+1)/3)}  
+\omega \frac{\eta'((\tau +2)/3)}{\eta((\tau+2)/3)}  \right) , \label{eq:Yi} \\ 
y_{\eta_3}(\tau) &=& \frac{-i}{\pi}\left( \frac{\eta'(\tau/3)}{\eta(\tau/3)}  +\omega\frac{\eta'((\tau +1)/3)}{\eta((\tau+1)/3)}  
+\omega^2 \frac{\eta'((\tau +2)/3)}{\eta((\tau+2)/3)}  \right)\,.
\nonumber
\end{eqnarray}
%
The overall coefficient in Eq. (\ref{eq:Yi}) is 
one possible choice; it cannot be uniquely determined. Thus we just impose the perturbative limit $y_{\eta_{1,2,3}}\lesssim\sqrt{4\pi}$
in the numerical analysis.
It implies that the mass hierarchy among the right-handed neutrinos could uniquely be fixed, therefore, one might say that DM candidate is determined by the structure of the modular function.
%
In the similar way as the Yukawa couplings, $M$ is also written by $y_{\eta_i}(k)$ such that
\begin{align}
\begin{pmatrix}M_{N_1}(\tau), M_{N_2}(\tau), M_{N_3}(\tau)\end{pmatrix}^T=
 M_0 \begin{pmatrix}
y_{\eta_1}(\tau)
,
y_{\eta_2}(\tau)
,
y_{\eta_3}(\tau)
\end{pmatrix}^T,
\end{align}
where $M_0$ can be taken as a free parameter determining the scale of right-handed neutrino mass.
Thus the mass relation among the three right-handed neutrinos are given once we fix modulus $\tau$.
The $A_4$ singlets couplings $y'_{\eta_i}$ with modular weight 4 are also written by 
\begin{align}
&\begin{pmatrix}y'_{\eta_1}(\tau), y'_{\eta_2}(\tau), y'_{\eta_3}(\tau)\end{pmatrix}_{1,1',1''} \nn\\
&=
\begin{pmatrix}
f_{1}^2(\tau) +f_{2}^2(\tau) +  f_{3}^2(\tau)
,
f_{1}^2(\tau) +\omega f_{2}^2(\tau) + \omega^2 f_{3}^2(\tau)
,
f_{1}^2(\tau) +\omega^2 f_{2}^2(\tau) + \omega f_{3}^2(\tau)
\end{pmatrix}_{1,1',1''}.
\end{align}
These structures are also determined by modulus $\tau$.

Higgs potential is given by
\begin{align}
{\cal V} &= -\mu_H^2 |H|^2 + {\mu_1 \mu_\varphi} |\varphi|^2 + \frac12{\mu_2 \mu_S} S^2  + \mu_{H\eta S} H^\dag\eta S
+\mu_{\varphi\eta} \varphi \eta^\dag\eta \nn\\
&+ \frac14 \lambda_H|H|^4 + \frac14 \lambda_\varphi|\varphi|^4 + \frac1{4!}\lambda_S S^4
+ \frac14\lambda_\eta |\eta|^4 
+ {\rm h.c.}
, \label{eq:pot}
\end{align}
where $\mu_H$, $\mu_1$, $\mu_2$, and $\lambda_H$ have zero modular weight.
Here $\varphi$ plays the role in inducing the mass of $\eta$ through $\mu_{\varphi\eta}$, after it develops a VEV $v_\varphi$.
Clearly, we can derive the inert conditions for $S$ and $\eta^0\equiv (\eta_R+i\eta_I)/\sqrt2$, and appropriate masses can be found.
In the CP even scalar sector, due to the preserved $Z_2$ symmetry, the inert scalar fields $S$ and $\eta_R$ doe not mix with the SM Higgs doublet but mix between themselves giving rise to the following physical inert scalar states:
\begin{align}
S=c_a H_1 + s_a H_2,\quad \eta_R =-s_a H_1 + c_a H_2,\quad s_{2a}=\frac{2v_H m_{H\eta S}}{m_{H_2}^2 - m_{H_1}^2},
\end{align}
where $s_a(c_a)$ is the short-hand symbol of $\sin a(\cos a)$.

After the electroweak spontaneous symmetry breaking,  the charged-lepton mass matrix is given by
\begin{align}
m_\ell&= \frac {v_H}{\sqrt{2}}
\left[\begin{array}{ccc}
y_e  & 0 & 0 \\ 
0 & y_\mu &  0 \\ 
0 & 0  & y_\tau  \\ 
\end{array}\right]\equiv
\left[\begin{array}{ccc}
m_e  & 0 & 0 \\ 
0 & m_\mu &  0 \\ 
0 & 0  & m_\tau  \\ 
\end{array}\right], 
\end{align}
while the right-handed neutrino mass matrix takes the form
\begin{align}
{\cal M_N} &=
\left[\begin{array}{ccc}
M'_{N_1}+M'_{N_2}+M'_{N_3} & M_{N_3} & M_{N_2} \\ 
M_{N_3} &M'_{N_1}+\omega^2 M'_{N_2}+\omega M'_{N_3}  & M_{N_1}   \\ 
M_{N_2} & M_{N_1} & M'_{N_1}+\omega M'_{N_2}+\omega^2 M'_{N_3}  \\ 
\end{array}\right],\label{eq:mn}
\end{align}
where $M'_{N_i}\equiv \zeta_i y'_{\eta_i} v_\varphi/\sqrt2$, ${\cal M_N}$ is diagonalized by $U_N{\cal M_N} U_N^T\equiv (M_{1},M_{2},M_{3})$, where $M_1\equiv M_X$ is expected to be the mass of DM candidate, $U_N$ is an unitary matrix, and free real parameter are $M_0$ and complex $\tau$.


{\it Lepton flavor violations} also come into our discussion and their formulas are given by~\cite{Baek:2016kud,Lindner:2016bgg}
\begin{align}
&{\rm BR}(\ell_i\to\ell_j\gamma)\approx\frac{48\pi^3\alpha_{em}C_{ij}}{G_F^2 m^2_{\ell_i}}(|a_{R_{ij}}|^2+|a_{L_{ij}}|^2),\\
&a_{R_{ij}}\approx\frac{m_{\ell_i}}{(4\pi)^2}\sum_{\alpha=1-3}Y^\dag_{j\alpha} Y_{\alpha i} F(M_{\alpha},m_{\eta^\pm}),\\
&F(m_a,m_b)\approx\frac{2 m^6_a+3m^4_am^2_b-6m^2_am^4_b+m^6_b+12m^4_am^2_b\ln\left(\frac{m_b}{m_a}\right)}{12(m^2_a-m^2_b)^4},
\end{align}
where $a_L=a_R(m_{\ell_i}\to m_{\ell_j})$, $C_{21}=1$, $C_{31}=0.1784$, $C_{32}=0.1736$, $\alpha_{em}(m_Z)=1/128.9$, $G_F=1.166\times10^{-5}$ GeV$^{-2}$, and $Y$ is written by
\begin{align}
& Y\equiv \lambda_3 y_{\eta_3} 
\left[\begin{array}{ccc} \hat\lambda_1& 0 & 0 \\ 0 & \hat\lambda_2 & 0   \\ 0 & 0 & 1   \\  \end{array}\right]
\left[\begin{array}{ccc} 1& 1 & 1 \\ 1 & \omega & \omega^2   \\ 1 & \omega^2 & \omega   \\  \end{array}\right]
\left[\begin{array}{ccc} \hat y_{\eta_1} & 0 & 0 \\ 0 &  \hat y_{\eta_2} & 0   \\ 0 & 0 & 1   \\  \end{array}\right] U^T_N,
\quad \hat y_{\eta_{1,2}}\equiv \frac{y_{\eta_{1,2}}}{y_{\eta_{3}}},\ \hat \lambda_{1,2}\equiv \frac{\lambda_{1,2}}{\lambda_{3}}.
\label{eq:Y}
\end{align}
The experimental upper bounds are given by~\cite{TheMEG:2016wtm, Aubert:2009ag,Renga:2018fpd}
\begin{align}
{\rm BR}(\mu\to e\gamma)\lesssim 4.2\times10^{-13},\quad 
{\rm BR}(\tau\to e\gamma)\lesssim 3.3\times10^{-8},\quad
{\rm BR}(\tau\to\mu\gamma)\lesssim 4.4\times10^{-8},\label{eq:lfvs-cond}
\end{align}
which will be imposed in our numerical calculation.
Notice here that muon anomalous magnetic moment is always obtained by {\it negative} value which is against the experimental value that tells us {\it positive} value. Thus, we will not discuss this issue furthermore.

 {\it Dark matter} candidate is expected to be the lightest fermion $X_R$ among three-right handed neutrinos and its mass is denoted by $M_X$.
The valid Lagrangian is given by
\begin{align}
\frac{Y_{i1}}{\sqrt2}\bar\nu_{L_i} X_R(\eta_R-i\eta_I) +Y_{i1} \bar\ell_{L_i} X_R \eta^-+{\rm h.c.},
\end{align}
Hereafter, we assume that the masses of $\eta_R,\eta_I,\eta^\pm$ are the same and symbolized by $m_\eta$ for simplicity.
Then the valid cross section to explain the relic density of DM is p-wave dominant in the expansion of relative velocity and its form is given by~\cite{Kubo:2006yx}
\begin{align}
\langle\sigma v_{rel}\rangle\approx \sum_{i,j=1-3}\frac{|Y_{i1}Y_{1j}^\dag|^2 r^2_X(1-2 r_X+2 r_X^2)}{24\pi M_X^2}v_{rel}^2,\quad r_X\equiv \frac{M^2_X}{m^2_0+M^2_X},
\end{align}
then one finds that the cross section should be within the following range to satisfy the observed relic density $\Omega h^2=0.1196\pm0.00031$~\cite{Aghanim:2018eyx};
\begin{align}
1.77552\lesssim {\langle\sigma v_{rel}\rangle}\times 10^9\ {\rm GeV^2} \lesssim1.96967,\label{eq:relic-cond}
\end{align}
where $v_{rel}^2\approx3/25$. 

\begin{figure}[tb]\begin{center}
\includegraphics[width=100mm]{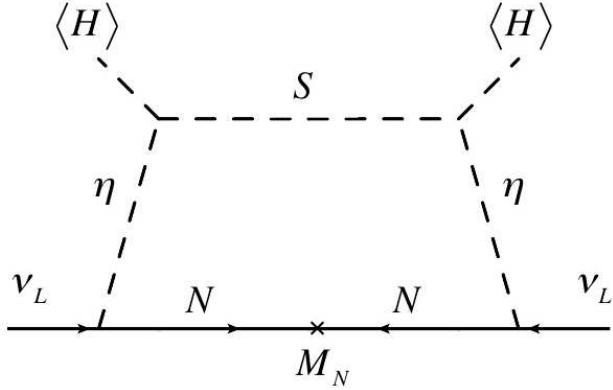}
\caption{One loop diagram generating neutrino mass.}   
\label{fig:diagram}\end{center}\end{figure}

{\it The light active neutrino mass matrix} is generated at one loop level as indicated by Fig.~\ref{fig:diagram} and is given by
\begin{align}
&m_{\nu_{ij}}\approx Y_{i\alpha} D_{N_\alpha} Y^T_{\alpha j},\\
&D_{N}\equiv \frac{\mu^2_{H\eta S} v^2_H}{4(4\pi)^2} 
\left[\begin{array}{ccc} M_1  F(m_{H_2},m_{H_1},M_1) & 0 & 0 \\ 0 &  M_2F(m_{H_2},m_{H_1},M_2) & 0   \\ 0 & 0 &  M_3  F(m_{H_2},m_{H_1},M_3)  \\  \end{array}\right],
\end{align}
with loop integration factor~\footnote{Once one successfully formulates a concrete form of $A_4$ triplet with 4 modular weight,
Ma-model~\cite{Ma:2006km} will be constructed on this framework.}
\begin{align}
F(m_1,m_2,m_3)\equiv & \frac{1}{(m_1^2-m_2^2)^2(m_1^2-m_3^2)^2(m_2^2-m_3^2)} 
\Biggl[ -(m_1^2-m_2^2)(m_2^2-m_3^2)(m_1^2-m_3^2) \nonumber \\
&  + 2(m_1^4+m_3^4)m_2^2\ln\left[\frac{m_1}{m_2}\right] 
 +2(m_1^4+m_2^4)m_3^2\ln\left[\frac{m_3}{m_1}\right]+ 4 m_1^2 m_2^2m_3^2\ln\left[\frac{m_2}{m_3}\right] \Biggr],
\end{align}
where 
we have assumed the mass insertion approximation; $m_{H_1}\approx m_S$ and $m_{H_2}\approx m_{\eta_R}$ ($s_a<<1$).
Then the neutrino mass matrix is diagonalized by an unitary matrix $U_{PMNS}$ as $U_{PMNS}m_\nu U^T_{PMNS}=$diag($m_{\nu_1},m_{\nu_2},m_{\nu_3}$)$\equiv D_\nu$, where $\sum D_{\nu}\lesssim 0.12$ eV is given by the recent cosmological data~\cite{Aghanim:2018eyx}.
Since the charged-lepton is mass eigenstate from the beginning, one identifies $U_\nu$ as $U_{PMNS}$.
Each of mixing is given in terms of the component of $U_{MNS}$ as follows:
\begin{align}
\sin^2\theta_{13}=|(U_{PMNS})_{13}|^2,\quad 
\sin^2\theta_{23}=\frac{|(U_{PMNS})_{23}|^2}{1-|(U_{PMNS})_{13}|^2},\quad 
\sin^2\theta_{12}=\frac{|(U_{PMNS})_{12}|^2}{1-|(U_{PMNS})_{13}|^2}.
\end{align}
Also, the effective mass for the neutrinoless double beta decay is given by
\begin{align}
m_{ee}=|D_{\nu_1} \cos^2\theta_{12} \cos^2\theta_{13}+D_{\nu_2} \sin^2\theta_{12} \cos^2\theta_{13}e^{i\alpha_{21}}
+D_{\nu_3} \sin^2\theta_{13}e^{i(\alpha_{31}-2\delta_{CP})}|,
\end{align}
where its observed value could be measured by KamLAND-Zen in future~\cite{KamLAND-Zen:2016pfg}.

\section{Numerical analysis}
In this section, we show numerical analysis to satisfy all of the constraints that we discussed above,
where we restrict ourselves the neutrino mass ordering is normal hierarchy.
First of all, we provide the allowed ranges for neutrino mixings and mass difference squares at 3$\sigma$ range~\cite{Esteban:2018azc} as follows:
\begin{align}
&\Delta m^2_{\rm atm}=[2.431-2.622]\times 10^{-3}\ {\rm eV}^2,\
\Delta m^2_{\rm sol}=[6.79-8.01]\times 10^{-5}\ {\rm eV}^2,\\
&\sin^2\theta_{13}=[0.02044-0.02437],\ 
\sin^2\theta_{23}=[0.428-0.624],\ 
\sin^2\theta_{12}=[0.275-0.350].\nn
\end{align}
The free dimensionless parameters $\kappa_i, \rho_i, \zeta_{i}^{L/R}, \epsilon_i$ (i=1-3) are taken to be the range of $[0.1-1]$,
while the mass parameters $M_0, m_{k}, m_\eta, m_{H_{1,2}}$ are $[1-5]$ TeV, where $m_\eta$ $m_k$ are respectively the masses of $\eta_0$ and $k^{\pm\pm}$.

\begin{figure}[tb]\begin{center}
\includegraphics[width=75mm]{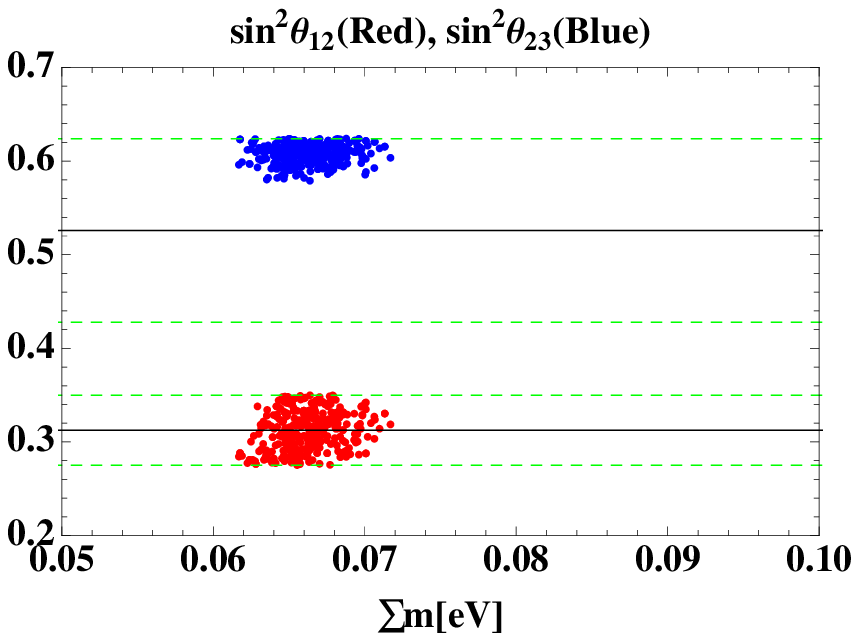}
\includegraphics[width=75mm]{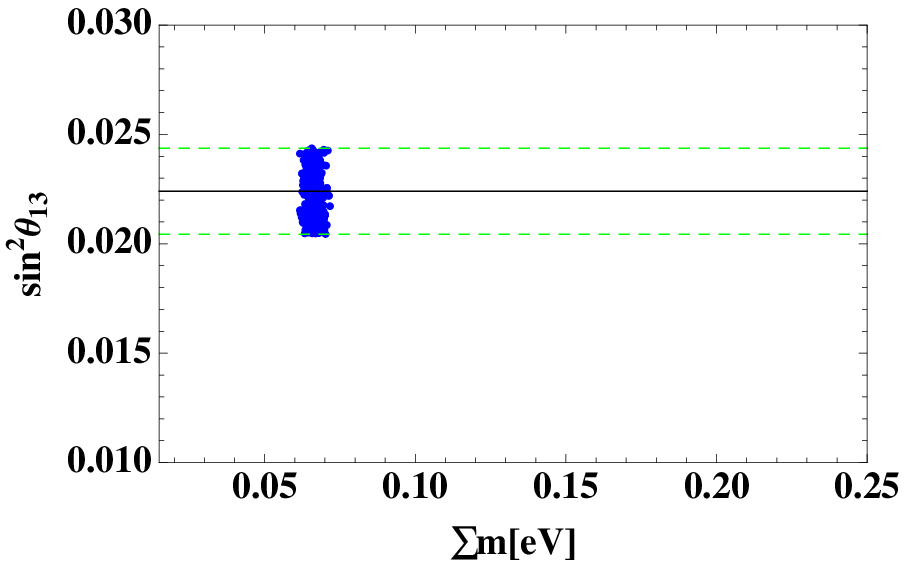}
\caption{The left side figure shows the sum of neutrino masses $\sum m(\equiv \sum D_\nu)$ versus $\sin^2\theta_{12}$(red color) and  $\sin^2\theta_{23}$(blue color), while
the right one demonstrates $\sum m$ versus  $\sin^2\theta_{13}$.}   
\label{fig:sins}\end{center}\end{figure}

The left side Fig.~\ref{fig:sins} shows the sum of neutrino masses $\sum m(\equiv \sum D_\nu)$ versus $\sin^2\theta_{12}$(red color) and  $\sin^2\theta_{23}$(blue color), while
the right one demonstrates $\sum m$ versus  $\sin^2\theta_{13}$.
Here, the black solid lines are the best fit values and the green dotted lines show 3$\sigma$ range. 
These figures suggest that the allowed region of sum of neutrino masses and $\sin^2\theta_{23}$ are respectively predicted to be around $0.062\lesssim\sum m\lesssim0.072$ eV and $0.56\lesssim\sin^2\theta_{23}\lesssim0.624$, while $\sin^2\theta_{12}$ and $\sin^2\theta_{13}$ lie whole over the experimental regions. One might be able to have an advantage of such a narrow region of $\sin^2\theta_{23}$ and be also discriminated from the other modular $A_4$ models.

\begin{figure}[tb]\begin{center}
\includegraphics[width=75mm]{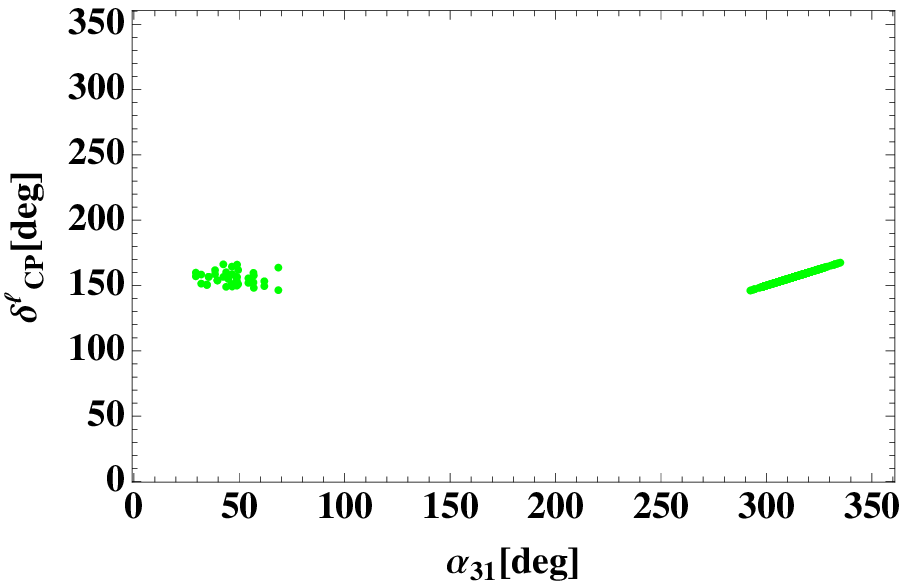}
\includegraphics[width=75mm]{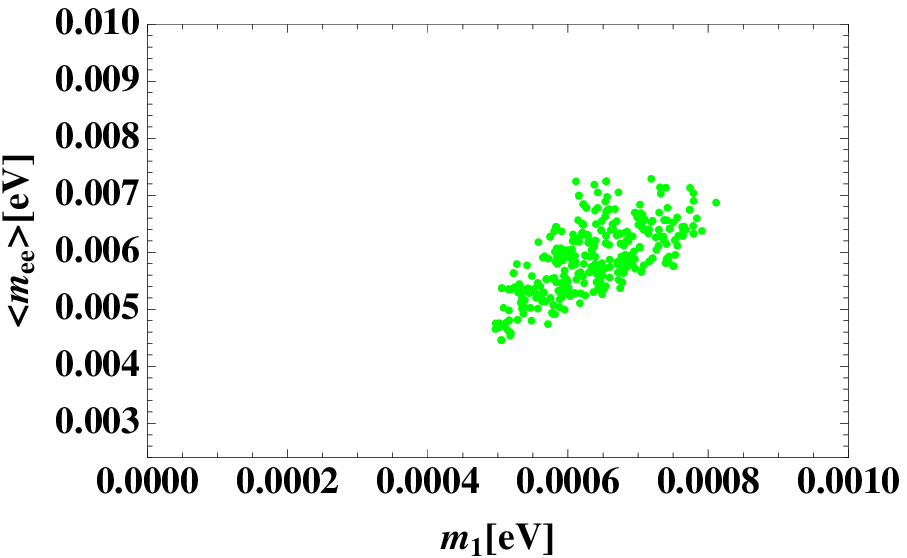}
\caption{The left side of figure shows Majorana phase $\alpha_{31}$ versus Dirac-CP phase $\delta^\ell_{CP}$, while
the right one demonstrates the first mass eigenstate $m_1$ versus neutrinoless double beta decay $\langle m_{ee}\rangle$}   
\label{fig:phase-mass}\end{center}\end{figure}

The left side of Fig.~\ref{fig:phase-mass} shows Majorana phase $\alpha_{31}$ versus Dirac-CP phase $\delta^\ell_{CP}$, while
the right side of Fig.~\ref{fig:phase-mass} demonstrates the first mass eigenstate $m_1$ versus neutrinoless double beta decay $\langle m_{ee}\rangle$.  
These figures suggest that $30^\circ\lesssim\alpha_{31}\lesssim 70^\circ$ or $290^\circ\lesssim\alpha_{31}\lesssim 340^\circ$,
$140^\circ\lesssim\delta^\ell_{CP}\lesssim170^\circ$, $5\times10^{-5}$eV$\lesssim m_1\lesssim8\times10^{-5}$eV, and 
0.0042eV$\lesssim \langle m_{ee}\rangle\lesssim$0.0072eV that are within the experimental bound.

\begin{figure}[tb]\begin{center}
\includegraphics[width=75mm]{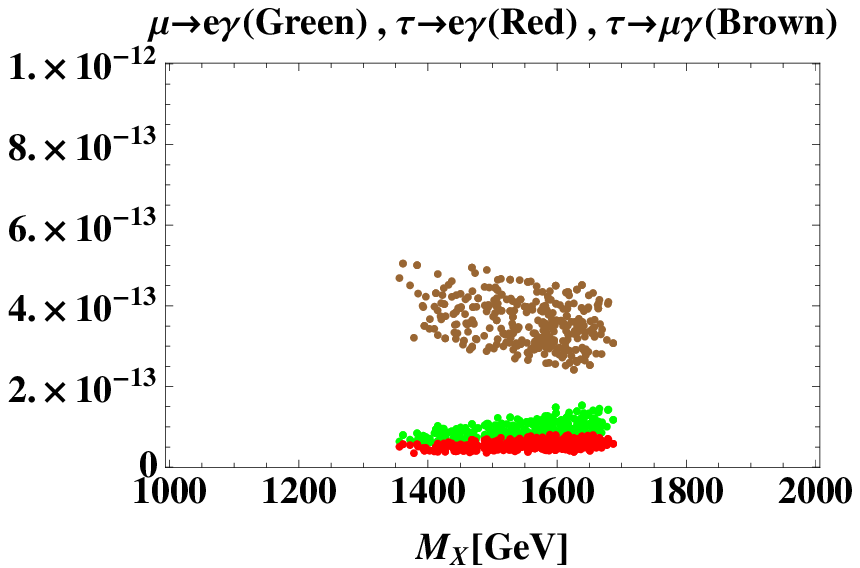}
\includegraphics[width=75mm]{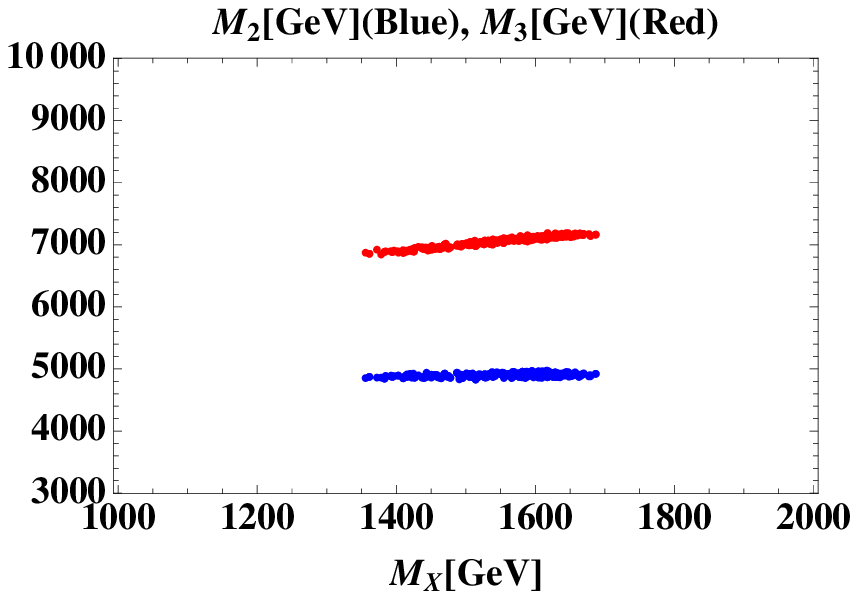}
\caption{The left side of figure shows the DM mass $M_X$ versus LFVs ($\mu\to e\gamma$(green),$\tau\to e\gamma$(red), and $\tau\to\mu\gamma$(brown)), while
the right one demonstrates $M_X$ versus the second mass $M_2(blue)$ and the third mass $M_3(red)$.
  }   
\label{fig:lfvs-M}\end{center}\end{figure}

The left side of Fig.~\ref{fig:lfvs-M} shows the DM mass $M_X$ versus LFVs; $\mu\to e\gamma$(green),$\tau\to e\gamma$(red), and $\tau\to\mu\gamma$(brown), while
the right side of Fig.~\ref{fig:lfvs-M} demonstrates $M_X$ versus the second mass $M_2(blue)$ and the third mass $M_3(red)$.
The figures suggest that all the LFVs are completely safe for the current experimental bounds, and
each of heavy Majorana fermion masses are uniquely fixed by 1.35TeV$\lesssim M_X\lesssim$1.7TeV, 4.8TeV$\lesssim M_2\lesssim$5.0TeV,
and 7.0TeV$\lesssim M_3\lesssim$7.2TeV.

Several remarks are in order:
 \begin{enumerate}
\item
The typical region of modulus $\tau$ is found in narrow space as 1.028\ $\lesssim\ $Re$[\tau]\lesssim$\ 1.103 and  1.219\ $\lesssim\ $Im$[\tau]\lesssim$\ 1.252.
\item $\alpha_{21}$ is found to be zero. 
\item 
Inverted ordering is numerically disfavored in our model.
\item 
Our cross section to explain the relic density via Yukawa sector is at most $5\times10^{-14}$ GeV$^{-2}$, which is much smaller than the correct one in Eq.(~\ref{eq:relic-cond}).
It implies that the relic density has to rely on the resonant effect which arises from s-channel cross section via additional Higgs $\varphi$~\cite{Kanemura:2010sh}. Therefore, the mass of $\varphi$ is at around the 2.7$\sim$3.4 TeV.~\footnote{Exactly speaking, $\varphi$ and $H$ mixes each other, but the mixing is restricted to be $\sim 0.2$. Thus, we can consider these $\varphi,H$ are almost mass eigenstate.}
%
 \end{enumerate}

\section{Conclusion and discussion}
\label{sec:conclusion}

We have proposed a modular $A_4$ symmetric model 
where the Majorana dark matter and the light active neutrino masses can be realized.
The stability of DM is assured by the modular weight, and the relation among DM and the other right-handed neutrino masses is characteristically determined by this symmetry; $M_X\ll M_2< M_3$,
once we embed the right-handed neutrinos into triplet under the $A_4$ symmetry and impose the constraints of lepton flavor violations and the relic density of DM. This feature could be in favor of realizing leptogenesis.
Unlikely to the typical flavor symmetries, the number of bosons are not many because the Yukawa couplings can also 
have an assignment under $A_4$ symmetry. As a result, the neutrino mass matrix is simply formulated at one-loop level and several predictions can be found as can be seen in the last part of section.

\section*{Acknowledgments}
\vspace{0.5cm}
{\it
The authors would like to thank Prof. Tatsuo Kobayashi for useful discussions and comments.
This research was supported by an appointment to the JRG Program at the APCTP through the Science and Technology Promotion Fund and Lottery Fund of the Korean Government. This was also supported by the Korean Local Governments - Gyeongsangbuk-do Province and Pohang City (H.O.). 
H. O. is sincerely grateful for the KIAS member, too.}


\end{document}